\documentclass[twoside,twocolumn,fleqn]{article}
\usepackage{epsf}
\usepackage{latexsym}
\usepackage{espcrc2}

\newcommand{\beq}[1]{\begin{equation}\label{#1}}
\newcommand{\eeq}{\end{equation}}
\newcommand{\bea}[1]{\begin{eqnarray}\label{#1}}
\newcommand{\eea}{\end{eqnarray}}
\newcommand{\rf}[1]{(\ref{#1})}
\newcommand{\trf}[1]{Table~\ref{#1}}

\newcommand{\vev}[1]{ {\langle #1 \rangle} }

\begin{document}
{\onecolumn

\title{Scaling and Quantum Geometry in 2d Gravity}

\author{K.~N.~Anagnostopoulos
\address{The Niels Bohr Institute,
Blegdamsvej 17, DK-2100 Copenhagen \O, Denmark}\\
\vspace{1mm}}

\begin{abstract}
We review the status of understanding of the fractal structure of the
quantum spacetime of 2d gravity coupled to conformal matter with $c\le
1$, with emphasis put on the results obtained last year. 
\end{abstract}

\maketitle}
\nopagebreak
\begin{table*}[hbt]
\setlength{\tabcolsep}{1.5pc}
\newlength{\digitwidth} \settowidth{\digitwidth}{\rm 0}
\catcode`?=\active \def?{\kern\digitwidth}
\caption{The fractal and spectral dimension of all $c\le1$ models studied.}
\label{t:1}
\begin{tabular*}{\textwidth}{@{}l@{\extracolsep{\fill}}rrrrrr}
\hline
\multicolumn{7}{c}{$d_h$}\\
\hline
\multicolumn{1}{l}{$c=-5$}        &
\multicolumn{1}{c}{$c=-2$}        &
\multicolumn{1}{c}{$c=0$}        &
\multicolumn{1}{c}{$c=1/2$}        &
\multicolumn{1}{c}{$c=4/5$}        &
\multicolumn{1}{c}{$c=1$}        &
\multicolumn{1}{c}{ } \\
\hline
3.236  & 3.562   &  4    & 4.21  & 4.42  & 4.83     & Eq.\protect\rf{*1}\\
1.236  & 2       &  4    & 6     & 10    & $\infty$ & Eq.\protect\rf{*2}\\
3.1-3.4& 3.58(4) & 4.0(1)& 4.1(1)& 4.0(1)& 4.1(3)   & Eq.\protect\rf{a:3}\\
       & 3.56(12)& 4.1(2)& 4.1(3)& 4.0(2)&          & Eq.\protect\rf{a:6}\\
       &         &       & 4.3(2)& 4.5(3)&          & Eq.\protect\rf{a:4}\\
\hline
\multicolumn{7}{c}{$d_s$}\\
\hline
       & 2.00(3) & 1.991(6) & 1.989(5) & 1.991(5) & &Eq.\protect\rf{a:5} \\ 
\hline
\end{tabular*}
\end{table*}

\section{INTRODUCTION}
The geometry of quantum spacetime of 2d gravity in the presence of
conformal matter with $c\le 1$ is the last important problem in those
models which is not yet fully understood. Although it is quite clear
\cite{transfer} that pure gravity ($c$=0) gives rise to a fractal
structure with Hausdorff dimension $d_h=4$ which becomes manifest by
the self similar distribution of geodesic boundary loop lengths at all
geodesic length scales, the
situation is not analytically understood 
in the presence of matter. In that case
the only rigorous tools available are numerical simulations. Several
critical exponents are defined in order to probe the geometry of
quantum spacetime. Among them are the fractal or Hausdorff dimension
$d_h$, the spectral dimension $d_s$ and the string susceptibility
$\gamma$. There exist scaling arguments
in the study of the diffusion of a fermion in the context of Liouville
theory which predict that \cite{ksw}
\beq{*1}
d_h=-2\frac{\alpha_1}{\alpha_{-1}}=
2\times\frac{\sqrt{25-c}+\sqrt{49-c}}{\sqrt{25-c}+\sqrt{1-c}}\, .
\eeq
In eq.~\rf{*1}, $\alpha_n$ denotes the gravitational dressing of a
$(n+1,n+1)$ primary spinless conformal field. In \cite{us,abt} a
remarkable agreement of numerical simulations with the above formula
was found for the non unitary $c=-2$ and $-5$ models. The situation is
less clear in the case of the unitary models $0<c\le 1$
\cite{syracuse,dif} where simulations seem to favour $d_h\approx
4$. At the present moment it cannot be resolved with certainty 
whether finite size effects
plague the results of the simulations, but it would be surprising if
this is the case since all critical exponents extracted from the
simulations (e.g. $\gamma$, scaling dimensions
of the fields) are in excellent agreement with the Liouville theory
results.

An alternative prediction for the fractal dimension comes from string
field theory \cite{deformation}
\beq{*2}
d_h=\frac{2}{|\gamma|}=\frac{24}{1-c+\sqrt{(25-c)(1-c)}}\, ,
\eeq
where a {\it modified definition of geodesic distance} has been
used. Such a prediction is in strong disagreement with simulations and
for a long time it was not understood whether the argument was wrong
or whether the simulations were not able to capture the correct
fractal structure due to the small size of the systems studied (notice
that for the unitary models with $1/2\le c\le 1$, $6\le d_h \le
\infty$). The results of \cite{us} pointed that there is a flaw in
\rf{*2} and recently in \cite{time} it has been suggested that one
cannot ignore the differences between the modified definition of
geodesic distance used in \rf{*2} and the real one and it is precisely
this difference that gives rise to \rf{*2}. The Ising model ($c=1/2$)
was studied in the loop gas representation and it was argued that the
distance used in \rf{*2} corresponds to absorbing the boundaries of
spin clusters to the geodesic boundary created by considering
successive spherical shells of increasing geodesic distance from a
given point. Simple mean field like scaling arguments for the
size of spin clusters, show that the increase of the volume of the
spherical shell scales non trivially with respect to the normal
definition of geodesic distance leading exactly to \rf{*2} for
$c=1/2$. The results are consistent with the performed numerical
simulations.

Last year progress has been made into understanding analytically the
spectral dimension $d_s$ of the above mentioned models
\cite{jw}. Ambj\o rn et. al. used a simple scaling hypothesis to
relate the spectral dimension $d_s$ to the {\it extrinsic} Haussdorf
dimension $D_h$ of the embedding of the corresponding bosonic
theory. They found that $1/d_s=1/D_h+1/2$ leading to $d_s=2$ for all
$c\le 1$. The basic scaling assumption made in the derivation, namely
the existence of well defined scaling dimensions for the diffusion
time and the geodesic distance for finite volume systems has been
numerically confirmed with great precision in \cite{dif}. 
Moreover, numerical simulations confirm that $d_s=2$ with high
precision \cite{syracuse,us,dif}. Notice that for the $c>1$,
$\gamma=1/2$  branched polymers all analytical and numerical
calculations give $d_s=4/3$.

\section{RESULTS}
The basic probe of the fractal structure of spacetime will be
correlation functions of the form
\begin{eqnarray}
\lefteqn{\vev{{\cal F}(\xi,\xi')}_{V,R}= 
  \int [{\cal D}g]Z_m[c,g]\delta(\int\sqrt{g}-V)}\nonumber\\
 & & \;\times\int d^2\xi d^2\xi'\sqrt{g}\sqrt{g'}{\cal F}(\xi,\xi')
     \delta(d_g(\xi,\xi')-R)\, ,\nonumber
\end{eqnarray}
which is a summation over all metrics modulo diffeomorphisms on a 2d
manifold of spherical topology and fixed volume $V$ weighted with the
partition function $Z_m[c,g]$ of the conformal matter fields of
central charge $c$, and we get contributions only from points
$\xi,\xi'$ separated by geodesic distance $d_g(\xi,\xi')=R$. In
particular one can define the volume of a spherical shell of geodesic
radius $R$ by $S_V(R)=\vev{1}_{V,R}/VZ_V$ ($Z_V$ is the fixed volume
partition function of the model), 2-point matter correlation functions
$S_V^\phi(R)=\vev{\phi(\xi)\phi(\xi')}_{V,R}/VZ_V$ and the probability
density of diffusing at distance $R$ after time $T$ $K_V(R,T)=
\vev{K_g(\xi,\xi';T)}_{V,R}/\vev{1}_{V,R}$. $K_g(\xi,\xi';T)$ is the
diffusion equation kernel defined by $\partial_T
K_g(\xi,\xi';T)=\Delta_g K_g(\xi,\xi';T)$ with
$K_g(\xi,\xi';0)=\delta(\xi,\xi')/\sqrt{g}$ and $\Delta_g$ being the
Laplacian of the metric $g$. From it one can define the return
probability $RP_g(T)=1/V\int\!\sqrt{g}\,K_g(\xi,\xi;T)$ and the
moments of the displacement $\vev{R^n(T)}_V=\int dR R^n\,
S_V(R)\,K_V(R,T)$. We expect the following
scaling relations to hold which can be used in the simulations in
order to compute $d_s$ and $d_h$
\bea{*4}
\label{a:3}
S_V(R)  = &V^{1-1/d_h} F_1(x) &\sim x^{d_h-1}\\
\label{a:4}
S_V^\phi(R) = & V^{1-1/d_h-\Delta} F\phi(x) &\sim x^{d_h-1-\Delta d_h}\\
\label{a:5}
RP_V(T) = & V^{-1} \Phi_0(y) &\sim y^{-d_s/2}\\
\label{a:6}
\vev{R^n(T)}_V=& V^{n/d_h}\Phi_n(y) &\sim y^{n d_s/2 d_h}\\
\label{a:7}
K_V(R,T)= & V^{-1} \Phi(x,y) & \, ,
\eea
where $x=R/V^{1/d_h}$, $y=T/V^{2/d_s}$ and the asymptotic relations
$\sim$ hold for $x\ll 1$ and $y\ll 1$. Remarkably, the simulations
show that the scaling relations are obeyed with excellent accuracy
over a wide range of distances $R$ and diffusion times $T$ even for
the small lattices (number of triangles $N>2000$) provided one uses
the simple finite size correction $R\to R+a$ where $a$ is the so
called ``shift'' (the shift in $T$ can be introduced as well but is
not as important) \cite{syracuse,us,abt,time}. 
The results for $d_h$
and $d_s$ are summarized in \trf{t:1}. One sees a clear agreement of
the $c<0$ models studied with \rf{*1}, whereas $d_h\approx 4$ for the
$0<c\le 1$ models. A notable exception are the 
\centerline{\epsfxsize=7.0cm \epsfysize=4.67cm 
 \epsfbox{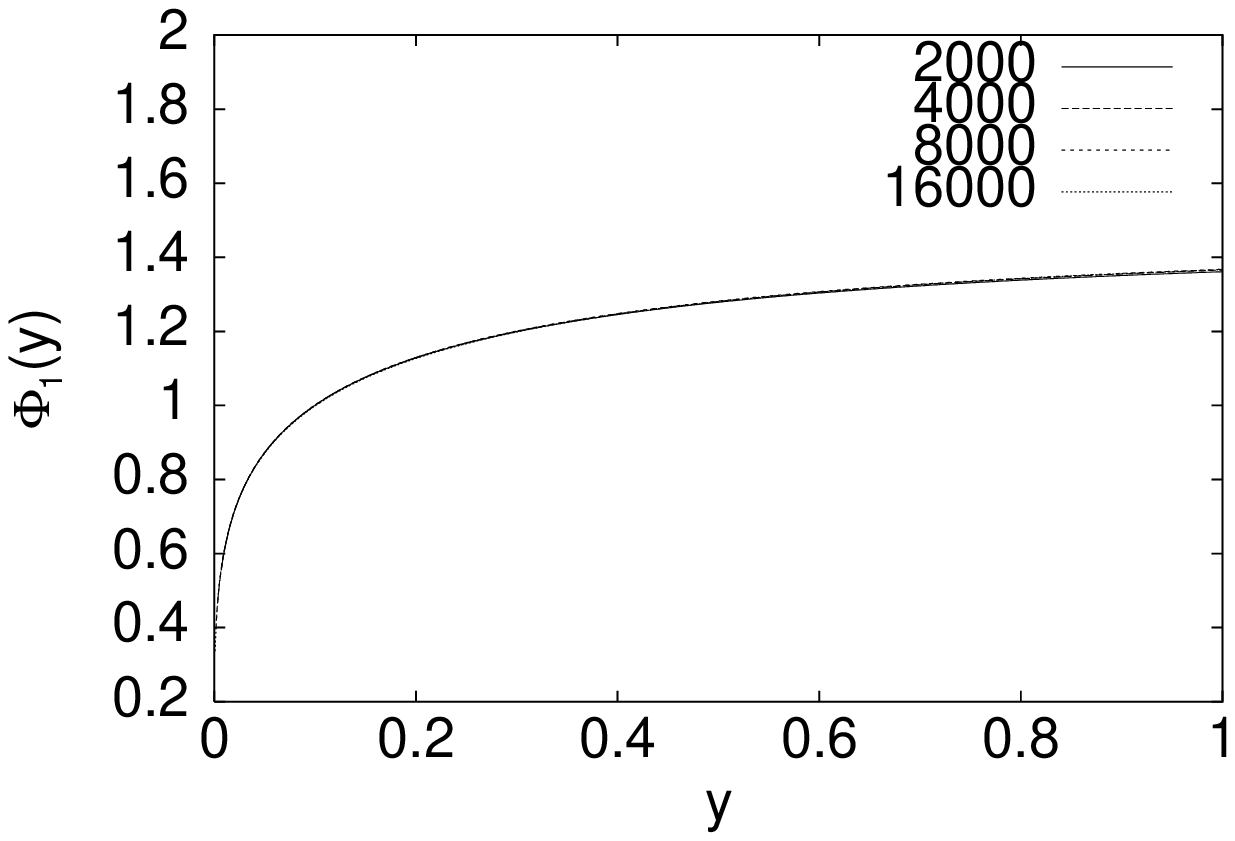}}

\vspace{-2mm}
\noindent {\bf Figure 1.} Collapse of $\vev{R^n(T)}_V$ according to
eq.~\protect\rf{a:6} for the Ising model.

\vspace{1mm}

\noindent
results obtained from
\rf{a:4} which seem to be not as inconsistent with the prediction of
\rf{*1}. 
But looking at the data more closely \cite{syracuse} one
observes that there are larger errors in determining $S_V^\phi(R)$ and
a small tendency of $d_h$ to decrease towards $d_h\approx 4$ with
volume. We also observe that the data clearly disagrees with \rf{*2}
for all models studied.
This is especially clear for the $c<0$ models
where the prediction of \rf{*2} gives a quite small value for $d_h$
and the lattices simulated have quite large linear sizes. In
\cite{time} the Ising model coupled to gravity was studied and
correlation functions $S_V(R')$ were computed where $R'$ is a modified
``geodesic distance'' corresponding to the discrete version of the one
used in the derivation of \rf{*2}: Given a set of triangles ${\cal
B}(R')$ at distance $R'$ from a given marked triangle, the
``geodesic'' boundary ${\cal B}(R'+1)$ at distance $R'+1$ contains all
triangles which share a link with ${\cal B}(R')$ which do not belong to
a ${\cal B}(R'')$ with $ R''\le R'$. In addition to those triangles, one
absorbs in ${\cal B}(R'+1)$ all triangles which belong to a boundary
of a spin cluster which crosses one of the triangles included in the
previous step. $R$ and $R'$ are not essentially different in the
magnetized phase of the model where a vanishing fraction of the
triangles of the lattice is crossed by a spin cluster boundary. In the
symmetric (dense) phase $R'$ is not well defined since almost the whole
lattice is crossed by spin cluster boundaries and $d_h'=\infty$. 
The numerical
simulations performed in \cite{time} measure $S_V(R')$ in the
pseudocritical region $\beta\to\beta_c^-$ and they observe that
$S_V(R')$ has the scaling properties \rf{a:3} with $d_h'\approx
5.0-5.8$ with  
\centerline{\epsfxsize=7.0cm \epsfysize=4.67cm 
 \epsfbox{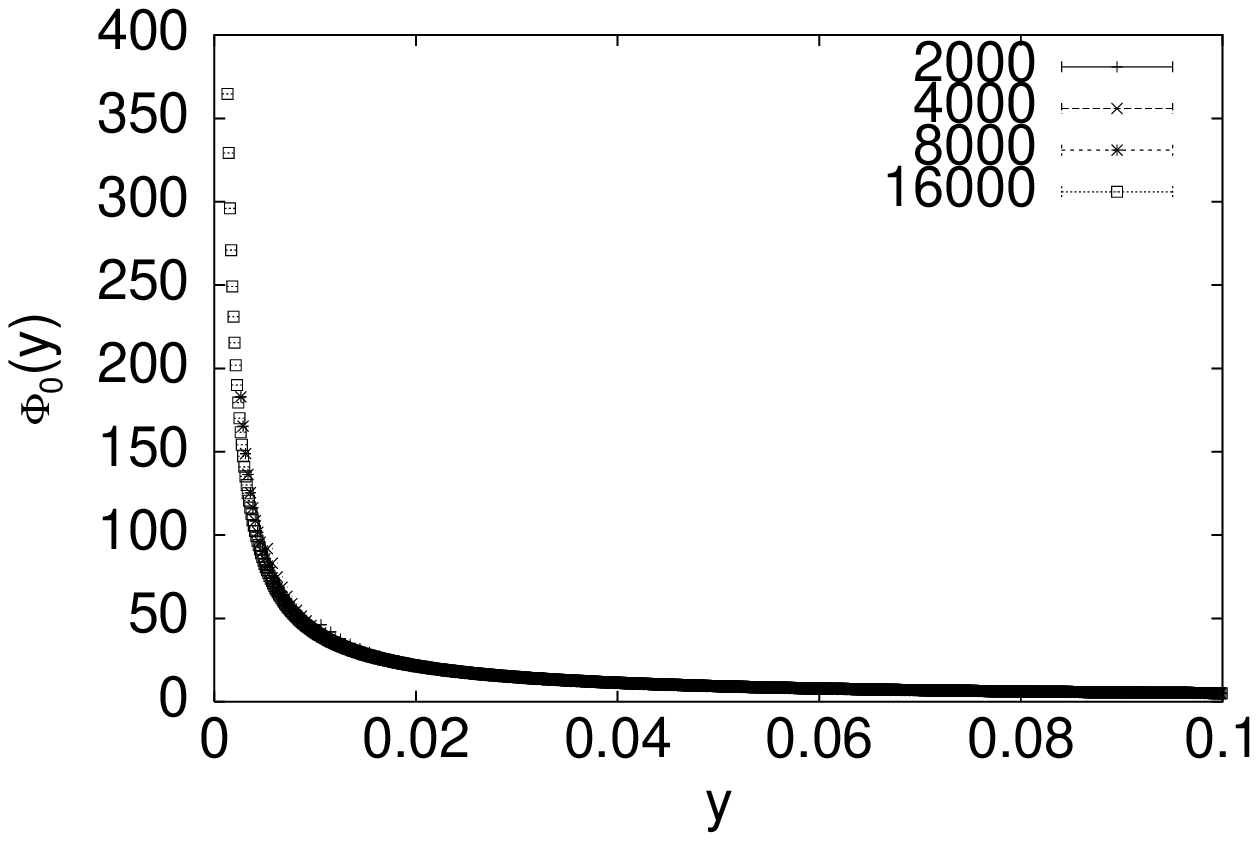}}

\vspace{-2mm}
\noindent {\bf Figure 2.} Collapse of $RP_V(T)$
according to eq.~\protect\rf{a:5} for the Ising model.

\vspace{1mm}

\noindent
a tendency of $d_h'$ to increase with lattice
size. $d_h'=6$,
given by \rf{*2}, is consistent with the numerical
data providing evidence that $R$ and $R'$ are essentially different
from each other and that $R'$ is the one that should be used in order
to realize \rf{*2}.

\end{document}